# On Orthogonalities in Matrices

R.N.Mohan[1]
Sir CRR Institute of Mathematics, Eluru-534007, AP, India
Email:mohan420914@yahoo.com

**Abstract:** In this paper we have discussed different possible orthogonalities in matrices, namely orthogonal, quasi-orthogonal, semi-orthogonal and non-orthogonal matrices including completely positive matrices, while giving some of their constructions besides studying some of their properties.

## 1. Introduction

There are various types of matrices in the literature having distinct properties useful for numerous applications, both practical and theoretical. The well-known matrix with orthogonal property is Hadamard Matrix, which was defined by Sylvester in 1867 [15] and was studied further by Hadamard in 1893 [7]. The Hadamard matrix H is the given square matrix satisfying $HH' = nI_n$, which has all the entries in the first row and the first column +1's and the rest of the elements are either +1's or -1's. The inner product of any two rows (columns) is 0. This is called as the orthogonal property. Hadmard conjectured that *the Hadamard matrix exists if and only if $n \equiv 0(mod 4)$*. Despite the efforts of several mathematicians, this conjecture remains unproved even though it is widely believed that it is true. This condition is necessary and the sufficiency part is still an open problem. These Hadamard matrices were systematically studied by Paley in 1933 [13]. There are some other orthogonal matrices like Slant, and Haar matrices, Discrete Fourier Transform, (DFT), Discrete Cosine transform (DCT), Jacket matrix and so on and for details refer to Moon Ho Lee and Hou Jia [12]. Then if the inner product is not zero then what are the other possibilities is a question for further exploration. Primarily there are four types of matrices namely i) orthogonal matrix, ii) quasi-orthogonal matrix, (two types), iii) semi-orthogonal matrix (two types), and iv) non-orthogonal matrix. In this paper we give some of the methods of constructions and study some of the properties of these four types of matrices.

The first type is the orthogonal matrix, which has the well-known example the Hadamard matrix. But in a given Hadamard matrix, if the elements of the principal diagonal are made zero then it ceases to be orthogonal i.e., some columns have orthogonal property and some other columns are not. Besides that it transcends into a trinary instead of a binary matrix. And these types of conference matrices are first studied by Belevitch [2], which are being used in conference telephony systems. He studied both the conference systems in relation with Hadamard system of matrices. These Hadamard matrices have received much attention in the recent past, owing to their well known and promising applications.

-----------------------------------

The orthogonal matrices have been studied by Moon Ho Lee and Hou Jia [12], besides giving many applications in communication systems, and Sarukhanyan, Agaian, and Astola in [14] studied the Hadamard matrices and their applications in image analysis, signal processing, fault-tolerant systems, analysis of stock market data, coding theory and cryptology, combinatorial designs and so on. While there is an abundance of literature depicting the importance of these orthogonal matrices still it is the natural inquisitiveness to know the status of non-orthogonal matrices also, which has been aimed in this paper.

## 2. Preliminaries

The **first type** is orthogonal matrix (real and complex). A real square matrix A is orthogonal if $A'A = AA' = I$. We consider real square matrices only for orthogonality as complex square matrices need not be orthogonal always.

The properties of orthogonal matrices are given by the following proposition refer to Abadir and Magnus [1].

**Proposition 2.1.** If A is an orthogonal matrix of order n, then (i) A is non-singular, (ii) $A' = A^{-1}$, (iii) $A'$ is orthogonal and (iv) if AB is orthogonal then B is also orthogonal.

For proof refer to [1]. ڤ

Now we have the following:

**Proposition 2.2.** If A is an orthogonal matrix then $\begin{bmatrix} A & A \\ A & -A \end{bmatrix}$ is also an orthogonal matrix and vice-versa. Proof is trivial. ڤ

The famous examples for orthogonal matrices are the Hadamard, Slant and Haar matrices, and Discrete Fourier Transform (DFT), Discrete Cosine Transform (DCT), and Jacket matrix as defined by Moon Ho Lee, for details refer to [12], who dealt with these matrices in it along with their applications.

**Note 2.1.** It may be noted that any matrix of order n × n, is said to be orthogonal if and only if $n \equiv 0 \pmod 4$. It can be seen from the formulae that have been obtained in [9, 10], among the three cases the orthogonal number is 0 means n=4k, which is true in the Type II and Type III when n is even. But they are not orthogonal. (Discussed in the later sections)

The **second type** is quasi-orthogonal matrix, which is defined as follows:

**Definition 2.1.** In this matrix the columns are divided into groups and the columns within each group are not orthogonal to each other but the columns in different groups are orthogonal to each other.

The conference matrix can be obtained by making the principal diagonal elements as zeros in a given matrix. Now considering an orthogonal matrix if the entries in the principal diagonal are made zero then this will give rise to quasi-orthogonal matrix, which is highly useful in tri-channel communication systems, with channels represented by 0's, -1's, and +1's.

**Example 2.1.** Consider the Hadamard matrix of order 4 as given by $\begin{bmatrix} 1 & 1 & 1 & 1 \\ 1 & -1 & 1 & -1 \\ 1 & 1 & -1 & -1 \\ 1 & -1 & -1 & 1 \end{bmatrix}$.

Let the diagonal elements be zero in this matrix. Then we get $\begin{bmatrix} 0 & 1 & 1 & 1 \\ 1 & 0 & -1 & 1 \\ 1 & 1 & 0 & -1 \\ 1 & -1 & -1 & 0 \end{bmatrix}$.

In this matrix, the columns 1 and 4 form as one group and the column 2 and 3 form as the second group. This is a quasi-orthogonal matrix having orthogonal numbers as 0 and 2.

This concept of quasi-orthogonality has been used in coding theory by Zafarkhani [8].

Alternatively, a new type of quasi-orthogonal matrix can be defined as follows:

**Definition. 2.2.** If the columns are divided into groups, the inner products of the columns of the same group are all zero, where as the inner products of the columns of different groups are either zero or some constant.

A method of construction of this second type is as follows: Consider a matrix of all +1's of order m×m, denoted by $\Uparrow$ and an orthogonal matrix of order n×n denoted by $H_{n \times n}$ and consider the Kronecker product of these two matrices, which is given by $\Uparrow_{m \times m} \otimes H_{n \times n}$. This gives a new pattern matrix of order mn×mn. In this resultant matrix, the columns are divided into m groups each has n columns. The inner products of the columns of the same group are all zeros and the inner products of different columns are either zero or mn.

**Example 2.2.** Consider $\Uparrow_{2 \times 2} \otimes H_{2 \times 2} = \begin{bmatrix} 1 & 1 \\ 1 & 1 \end{bmatrix} \otimes \begin{bmatrix} 1 & 1 \\ 1 & -1 \end{bmatrix} = \begin{bmatrix} 1 & 1 & 1 & 1 \\ 1 & -1 & 1 & -1 \\ 1 & 1 & 1 & 1 \\ 1 & -1 & 1 & -1 \end{bmatrix}_{4 \times 4}$, where as it has two orthogonal numbers 0 and 4. Hence it is a quasi-orthogonal matrix. The columns have been divided into two groups (1, 3) and (2, 4).

This represents a signal, where as in some intervals it is running in certain pattern but when compared to one interval to another interval the signal behavior is uneven and turbulent. These types of matrices are also useful in the construction of codes.

Now referring to Abdir and Magnus [1] we have the following:

**Proposition 2.3.** Let A and B be square matrices of order n, and then $AB = I_n$ if and only if $BA = I_n$.

**Proof.** Let A and B be matrices of order n. Since $AB = I_n$, we have $N = rk(AB) \leq \min(rk(A), rk(B)) \leq rk(A) \leq n$. Hence, A (and B) has full rank n, and is therefore nonsingular. And since the inverse is unique, $BA = I$ holds good. Hence the result. ف

**Proposition 2.4.** Let A be a matrix of order m×n. If m < n, then there exists no m×n matrix B such that $B'A = I_n$.

**Proof**. If $B'A = I_n$, then $n = rk(I_n) = rk(B'A) \leq rk(A) \leq m$, which is a contradiction. Hence the result. ف

The **third type** is called as semi-orthogonal matrix, which is defined by Abadir and Magnus in [1], as follows:

**Definition 2.3.** If A is a real m × n matrix, where m ≠ n, such that $AA' = I_m$ or $A'A = I_n$, but not necessarily both, is called semi-orthogonal matrix.

As an example for this type of semi-orthogonal matrix, let x be an n ×1 vector (n>1) such that $x'x = 1$, for example x: = $e_i$. Le A := x' (and hence m = 1), Then $AA' = x'x \neq I_n$.

This can be considered as the first type of semi-orthogonal matrix.

And then the second type of semi-orthogonal matrix, which is a new pattern matrix that can be defined as follows:

**Definition 2.4.** In the given n×2n real matrix from the 2n columns it has, the first n columns are orthogonal to each other and the remaining n columns are not orthogonal to each other. Besides in the two groups any column of the first group is not necessarily orthogonal with any column of the second group.

The method of construction is as follows: By juxtaposing an orthogonal matrix of order n × n with a non-orthogonal matrix of same order n×n will give a semi-orthogonal matrix of order n × 2n as defined above.

**Note 2.2.** While associating with the Hadamard matrices, the M-Matrices of Type I or III when n is even and of same order should alone be taken.

**Example 2.3.** Consider an orthogonal Matrix H and a non-orthogonal matrix M, and by adjoining these two matrices, we get a semi-orthogonal matrix, with orthogonal numbers -2, 0, 2.

$$H = \begin{bmatrix} 1 & 1 \\ 1 & -1 \end{bmatrix}_{2 \times 2}, \quad M = \begin{bmatrix} 1 & -1 \\ -1 & 1 \end{bmatrix}_{2 \times 2} \quad \text{Then} \quad HM = \begin{bmatrix} 1 & 1 & 1 & -1 \\ 1 & -1 & -1 & 1 \end{bmatrix}_{2 \times 4}.$$

There is subtle difference between the quasi-orthogonal and semi-orthogonal matrices.

The concept of semi-orthogonalilty has been used by Groß, Trenkler, and Troschke in [6].

This second type of semi-orthogonal matrices are useful in signal processing, where as the signal under consideration is flowing without any significant changes to some extent and suddenly starts flow with turbulent fluctuations like that showing the cardiac functioning, as depicted by an electrocardiogram.

And the **fourth type** is non-orthogonal matrix in which no two columns of the matrix are orthogonal. In this case we have to evaluate orthogonal numbers and their properties. The good examples for non-orthogonal matrices are given by Mohan, Kageyama, Lee, and Yang [9], and Mohan, Lee, and Paudel [10], in which some non-orthogonal matrices called M-matrices of Type I, II, and III have been defined and constructed and their properties have been studied.

It should not be taken like haphazardly filling the entries by +1 and -1 simply gives a non-orthogonal matrix. Because the accountability of orthogonal numbers and the number of +1's and -1's in each row and each column becomes perilously complex. But the constructions of these non-orthogonal matrices, avoiding these two difficulties have been given in Mohan et al [9, 10], while studying the pattern of these matrices.

In fact all these matrices are having practical applications in one way or the other and also have specific properties, which we will study besides giving some new patterns of these matrices.

The binary matrices are those having only two types of entries, either (0, 1), (1,-1) or literally any two types of entries. But usually we consider the above two types only in our present discussions. The binary matrices were first studied by Ehlich [4], and Ehlich and Zeller [5].

If the matrix A can be decomposed as A = BB$^T$, then the real matrix A is called positive semi-definite. In some applications the matrix B has to be element-wise non-negative. If such a matrix exists then A is called completely positive. This is a non-orthogonal matrix. But if B has entries ±1's such that BB$^T$ = nI$_n$ instead of A, then B is an orthogonal matrix. The completely positive matrices appear in the study of block designs in combinatorics, probability, and in various applications of statistics, including a Markovian model for DNA evolution and a model for energy demands. These types of matrices and their applications were discussed by Berman and Manderer in [3].

## 3. Some main properties

A *square* matrix M is called *orthogonal* if its inverse $M^{-1}$ equals to its transpose $M^T$, this means that

$$M \cdot M^T = I$$

Observe that this implies, in particular, that the determinant det (M) = ±1

Orthogonal matrices have a number of useful properties. Columns of an orthogonal matrix treated as vectors make out an *ortho-normal basis* of the space (this means that they are pairwise orthogonal and each one is of unit length). The same of course holds good for rows, as well. Conversely, having an ortho-normal basis and putting the vectors side by side, makes an orthogonal matrix. Will this be possible in the case of the other types of matrices, for which answer can not be affirmative. The columns of the other types of matrices do not form an ortho-normal basis.

Two ortho-normal bases are said to be *compatibly oriented* if the (orthogonal) matrix representing the change of variables from one to the other has determinant +1. This happens if and only if one basis can be rotated onto the other.

The linear transformation corresponding to an orthogonal matrix is called a (linear) *isometry* (or a *rigid (linear) transform*). If the determinant of the matrix equals +1 then it is a rotation, if it equals −1 it is a reflection. Orthogonal matrices of given dimension n form a group, called as an *orthogonal group* and is denoted by O (n). The orthogonal matrices of determinant 1 (i.e. those associated with rotations) form a subgroup called as a *special orthogonal group* and is denoted by SO (n).

An isometry preserves many geometric properties of the space. In particular it preserves lengths, perpendicularity and angles. Since the inverse and transpose of the matrix of the linear isometry cancel out, normals are treated by an isometry in the same way as ordinary vectors. This is not so for a general transformation.

If we consider the other types of matrices viz., quasi-orthogonal matrix, semi-orthogonal matrix and non-orthogonal matrix we can not form the ortho-normal basis, isometry, and hence we can not say how the normal behaves, but it can be explored further.

And the nature of non-orthogonal matrix has yet to be studied as above. But Mohan et al in [9, 10] defined three types of non-orthogonal matrices called M-matrices of Type I, II, and III, and have studied their properties. When the matrix is non-orthogonal then it will have orthogonal numbers. In the three cases it has been formulated. The sum of these orthogonal numbers and some of their properties also have been studied.

Now in the case of orthogonal matrix the orthogonal number is zero. And in the case of the other matrices we have to formulate the orthogonal numbers and the orthogonal number that has been defined in [9, 10] is as follows:

**Definition 3.1.** The **orthogonal number** of a given matrix with entries ± 1 is defined as sum of the products of the corresponding entries in the two given rows of the matrix considered, which is called as inner product of the rows. Consider any two rows $R_l = (r_1, r_2, ..., r_n)$ and $R_m = (s_1 s_2, ..., s_n)$ and then the orthogonal number denoted by $g_i$ can be defined as $g_i = (R_l R_m) = \sum_{j=1}^{n} r_j s_j$.

In the case of the Hadamard matrix as it is orthogonal matrix then all the orthogonal numbers $g_i$'s are equal to zero. The Hadamard matrix is defined as a square matrix with entries ±1, such that HH′ = $nI_n$. And the Hadamard conjecture states that the Hadamard matrix exists if and only if n ≡ 0 (mod 4). (Of course for n =2 also the Hadamard matrix exists). It is only necessary condition and for n = 4k, there exists numerous binary matrices with entries ± 1, which are not orthogonal matrices. Let us consider a matrix M of order n × n, where n is an integer, with entries +1 or -1 and evaluate the inner product of any two rows, which we call as an orthogonal number as defined earlier and is denoted by $g_i$.

The non-orthogonal pattern matrices have been constructed by considering three possible ways by taking $a_{ij} = (d_i \otimes d_h d_j) \bmod n$ and by suitably defining $d_i, d_h, d_j, \otimes$ and n as follows:

1. $a_{ij} = 1 + (i-1)(j-1) \bmod n$, when n is prime.

2. $a_{ij} = (i.j) \bmod n$, when n+1 is prime.
   (We considered n+1 as a prime for our need here).

3. $a_{ij} = (i+j) \bmod n$, when n is a positive integer.

By these three types we get M-matrices of Type I, II, and III and the formulae for the orthogonal numbers are given by i) g = 4k+2-n, ii) g = 4k-n, since n+1 is prime, n is even in this case, and iii) g = 4k-2-n, when n is odd and when is even it is same as in type II and k is a positive integer with a restriction and for details refer to [9,10]. These three types are combinatorially equivalent but structurally different and hence studied distinctively. Observing these formulae, we can trivially know that if g is zero then n = 4k, which is true in the case of Type II and in Type III when n is even. Hence in the Hadamard matrix n≡0(mod 4) is holding good as a necessary condition. For being sufficiency part there will be numerous matrices of order 4k, which need not be orthogonal, for example the M-matrices of Type II and M-matrix of Type III when n is even. For more details refer to [9, 10].

The Hadamard Matrix H is a square n × n matrix, with entries ±1, and the first row and the first column consists of +1 only, such that HH′ = $nI_n$. It is orthogonal matrix, where the inner product of any two rows denoted by $g_i$ is 0.

The Hadamard conjecture states that *The Hadamard matrix of order n exists if and only if n≡0 (mod4). That is, n is a multiple of 4. Let n = 4k, where k is an integer. We have to establish that if g = 0, n = 4k and conversely. Then it is a Hadamard matrix as mentioned above.*

This has been considered in Mohan [11] and Tressler [16] and can be referred to.

In fact we have the following result.

**Result 3.1.** In an orthogonal matrix of order n×n there are n rows, and n columns, each set of which can be permutated among them in n! ways. So the total number of arrangements will be $(n!)^2$. It is the possible number of orthogonal matrices that can be generated from the given orthogonal matrix of order n.

**4. The methods of constructions**

We give hereunder the methods of constructions of various new types of matrices obtained by making use of orthogonal and non-orthogonal matrices.

Let M be a non-orthogonal matrix of one of the M-matrices of Type I, II, and III, and H be an orthogonal matrix and $\Uparrow$ be a matrix of all ones. Let $g_i$ denotes the orthogonal number of the M-matrix of Type I, II, or III given by the formulae $g_i$ = 4k+2-n, 4k-n, or 4k-2-n respectively, for details refer to [9,10].

**Note 4.1.** It can be visualized that when an orthogonal matrix and a non-orthogonal matrix act as parents and that all the quasi-orthogonal and semi-orthogonal matrices are their offspring. The cross fertilization of them will lead to some new types of matrices which will be of interest, both theoretical and application points of view.

Now see the following patterns, starting with some trivial constructions.

**Proposition 4.1.** The matrices given by $\Uparrow_{m \times m} \otimes H_{n \times n}$, $\Uparrow_{m \times m} \otimes M_{n \times n}$, are higher order non-orthogonal matrices, which have the orthogonal numbers in the first case 0 and mn and in the second case mn and $mg_i$'s.

**Proof.** In the case of $\Uparrow_{m \times m} \otimes H_{n \times n}$, it is a quasi-orthogonal matrix with orthogonal numbers mn and 0. And in the case of $\Uparrow_{m \times m} \otimes M_{n \times n}$, if the orthogonal numbers of M are $g_i$'s (whatever may the Type I, II, or III), the orthogonal numbers of the resultant matrix are mn and $mg_i$'s. ڤ

In some of the propositions below there is an amalgamation of the concepts of orthogonality and non-orthogonality, which gives rise to some new types of matrices, where both orthogonality and non-orthgonolity play their role.

**Proposition 4.2.** The matrices given by $[M:H]$, $[H:M]$, $[M \otimes H]$, $[H \otimes M]$ are having the orthogonal numbers depending on the type of M-matrix, but both M and H should be of same order. Here (:) indicates juxtaposition and ($\otimes$) indicates Kronecker product of matrices.

**Proof.** In both of the $[M:H]$, $[H:M]$ structures the orthogonal numbers are 0 and $4g_i$'s.

And in the structures $[M \otimes H]$, the orthogonal numbers of the resultant matrix can be obtained as follows: Any two rows of the same block the orthogonal number is zero and any two rows of different blocks of the resultant matrix the orthogonal numbers are given by 4$g_i$ or zero.

Alternatively, if we take the pattern $[H \otimes M]$, the orthogonal numbers of any two rows of the same block are given by 4$g_i$'s and any rows of different blocks are given by 0. ف

**Proposition 4.3.** The pattern [H: H: …: H], juxtaposing for m times of the orthogonal matrix H gives a quasi orthogonal matrix.

**Proof.** The $H_{n \times n}$ and juxtaposing for m times adjacently gives another quasi-orthogonal matrix with orthogonal numbers $g_i$'s as 0 and n. The resultant matrix is of order n×mn. ف

**Proposition 4.4.** Let D be the matrix obtained from [H: H: …: H], then $DD^T$ = mn$I_n$ and $D^T D = \underline{\Uparrow}_{m \times m} \otimes nI_n$.

**Proof.** Let D be the matrix obtained by the above pattern order n×mn. The order of $D^T$ is mn ×n. Hence if we consider $DD^T$ is of order n×n and the order of $D^T D$ is mn×mn and trivially for the first type $DD^T = mn\underline{\Uparrow}_{n \times n}$, and for the second type $D^T D = \underline{\Uparrow}_{m \times m} \otimes nI_n$. And the orthogonal numbers can be easily calculated, and this is also a quasi orthogonal matrix. ف

**Proposition 4.5.** The pattern $\begin{bmatrix} M & M \\ M & -M \end{bmatrix}$' yields quasi-orthogonal matrices and the orthogonal numbers are given by 2$g_i$ and 0.

**Proof.** Let the orthogonal number of M is given by $g_i$, then consider any two rows of the same block (M M) or (M -M), the orthogonal numbers are given by 2$g_i$. And any two rows one from one block and the other from the other block is zero. As a further extension of this result we can have the following proposition. ف

**Proposition 4.6.** The orthogonal numbers for the following pattern are $g_i = 0$, and 4$g_i$.

$$\begin{bmatrix} M & M & M & M \\ M & -M & M & -M \\ M & M & -M & -M \\ M & -M & -M & M \end{bmatrix}.$$

**Proof.** There are four horizontal blocks. Consider the first block and orthogonal number of any two rows in it is 4$g_i$. Similarly in the other blocks the orthogonal numbers of any two rows is again 4$g_i$. The orthogonal number of any two different rows of distinct blocks is zero. Thus this pattern gives a quasi-orthogonal matrix having orthogonal numbers depending on the type of the matrix, which may be any one of the types, I, II, or III, besides having the inner product of some distinct rows as 0.

**Note 4.2**. In the similar way this can be further generalized.

**Proposition 4.7.** The matrices given by

$$\begin{bmatrix} H & M \\ M & -H \end{bmatrix}, \begin{bmatrix} M & H \\ H & -M \end{bmatrix}$$ give some quasi- or semi- orthogonal matrices.

Proof. Let H and M be the matrices of the same order, each of their rows the number of +1's and the number of -1's be equal to n/2. If two distinct rows of the same block are considered it will be $g_i$'s only. Two different rows of distinct blocks are considered the orthogonal numbers will be 0 or $g_i$'s. In this case the orthogonal number is given by 2(4k-n).    ڤ

**Proposition 4.8.** Consider an orthogonal matrix H of order n, and the rows of this matrix be permuted as $R_i$ to $R_{i+1 (\text{mod } n)}$. Then each matrix thus obtained is numbered as $H_1, H_2, ..., H_n$. Let us consider M-matrix of Type II. In the place of each entry of this matrix let there be the corresponding matrix as mentioned above. Then we obtain a quasi-orthogonal matrix with orthogonal numbers 0 and n.

**Proof.** The M-matrix of Type II is given by
$$\begin{bmatrix} 1 & 2 & ... & n \\ 2 & 4 & ... & n-1 \\ ... & ... & ... & \\ n & n-1 & ... & 1 \end{bmatrix}.$$ Hence by replacing each entry by the corresponding H-matrix,

where i = 1, 2, …, n indicates the permutation. Then we have

$$\begin{bmatrix} H_1 & H_2 & ... & H_n \\ H_2 & H_4 & ... & H_{n-1} \\ ... & ... & & .. \\ H_n & H_{n-1} & ..... & H_1 \end{bmatrix}.$$

This is a quasi-orthogonal matrix with orthogonal numbers 0 and n, by considering the inner products of the rows of the horizontal blocks.

**Example 4.1.** The M-matrix of type II is given by $\begin{bmatrix} 1 & 2 & 3 & 4 \\ 2 & 4 & 1 & 3 \\ 3 & 1 & 4 & 2 \\ 4 & 3 & 2 & 1 \end{bmatrix}$. Hence by replacing each entry by the corresponding H-matrix, in which the subscript i = 1,2,3,4 indicates the permutation. Then we have

$$\begin{bmatrix} H_1 & H_2 & H_3 & H_4 \\ H_2 & H_4 & H_1 & H_3 \\ H_3 & H_1 & H_4 & H_2 \\ H_4 & H_3 & H_2 & H_1 \end{bmatrix}.$$

**Proof.** This gives a quasi–orthogonal matrix having orthogonal numbers 0 and 4. Any two distinct rows of any block have orthogonal number 0 and any two distinct rows of different blocks, also have orthogonal number 0, but if the two rows, there will be one entry in a vertical block coincidence then the orthogonal number is 4. ڤ

**Proposition 4.9.** Consider an orthogonal matrix H of order n, and the rows of this matrix be permuted as $R_i$ to $R_{i+1(\mod n)}$. Then let each matrix thus obtained be numbered as $H_1, H_2, ..., H_n$. Let us consider a circulant matrix of order n with the list 1, 2, …, n. In the place of each entry of this circulant matrix let there be the corresponding matrix. Then we obtain a quasi-orthogonal matrix with orthogonal numbers 0 and n.

**Proof.** The circulant matrix with the list 1, 2, …, n  is given by

$$\begin{bmatrix} 1 & 2 & ... & n \\ 2 & 3 & ... & n-1 \\ ... & ... & ... \\ n & n-1 & ... & 1 \end{bmatrix}.$$ Hence by replacing each entry in it by the corresponding H-matrix as above,

we get the resultant pattern as follows:

$$\begin{bmatrix} H_1 & H_2 & ... & H_n \\ H_2 & H_3 & ... & H_1 \\ ... & ... & & ... \\ H_n & H_1 & ... & H_{n-1} \end{bmatrix}.$$

**Proof.**  Now considering the horizontal blocks in the structure and the rows in it and we can easily have the orthogonal numbers as 0 and $n^2$.

**Example 4.2.**  The circulant matrix with the list 1, 2, 3, 4 is given by

$$\begin{bmatrix} 1 & 2 & 3 & 4 \\ 2 & 3 & 4 & 1 \\ 3 & 4 & 1 & 2 \\ 4 & 1 & 2 & 3 \end{bmatrix}.$$ Hence by replacing each entry in it by the corresponding H-matrix as above, we get the resultant pattern as follows:

$$\begin{bmatrix} H_1 & H_2 & H_3 & H_4 \\ H_2 & H_3 & H_4 & H_1 \\ H_3 & H_4 & H_1 & H_2 \\ H_4 & H_1 & H_2 & H_3 \end{bmatrix}.$$

**Proof.** This gives a quasi–orthogonal matrix having orthogonal numbers 0 and 16. Any two distinct rows of any block have orthogonal number 0 and any two distinct rows of different blocks, also have orthogonal number 0, but if the two rows coincide then the orthogonal number is 16. This can be further generalized and the orthogonal numbers can be easily obtained in the same way. ڤ

**Proposition 4.10.** Consider a non-orthogonal matrix of order n, and the rows of this matrix be permuted as $R_i$ to $R_{i+1(\mod n)}$. Then each matrix thus obtained be numbered as $M_1, M_2, ..., M_n$, and consider a circulant matrix of order n with the list 1, 2, …, n. In the place of each entry of this circulant matrix let there be the corresponding matrix. Then we obtain a quasi-orthogonal matrix with orthogonal numbers 0 and n, the matrix pattern is as follows:

$$\begin{bmatrix} M_1 & M_2 & M_3 & M_4 \\ M_2 & M_3 & M_4 & M_1 \\ M_3 & M_4 & M_1 & M_2 \\ M_4 & M_1 & M_2 & M_3 \end{bmatrix}.$$

**Proof.** In this pattern, consider any two rows of the same block the orthogonal numbers are $4g_i$. Now if we consider distinct blocks and from them two different rows one from each block the orthogonal numbers will be again $4g_i$'s may be different to the above. This can be further generalized and the orthogonal numbers can be easily obtained in the same way.

**Note 4.3.** Here also we can as well take the matrix of Type II or otherwise, and can obtain some new matrix.

There are numerous other ways of generating some new matrices by many other distinct methods, leading to a new evolution of matrices as above, and many of the algebraic properties are still open, which are worth tractable.

Further work in this direction can be seen in a sequel to this paper to appear shortly.

**Acknowledgements:** The author expresses his deep sense of gratitude to Prof. M.G.K.Menon, who is a constant source of inspiration for furthering his research and the authorities of Sir CRR Institutions Management Committee, Secretary G.Subbarao, AO Sri K.Srimannarayana, and Principal R. Surya Rao. Also his thanks are due to Prof. Bill Chen, Director Center for Combinatorics, and Vice-President, Nankai University, Tianjin-300071, PR China and the Third World Academy of Sciences, Trieste, Italy for facilitating him to further his research studies.

\